\documentclass[%
 reprint,
 amsmath,
 amssymb,
 aps,
 prd,
]{revtex4-2}

\usepackage{graphicx}
\usepackage{dcolumn}
\usepackage{bm}
\usepackage{hyperref}

\begin{document}

\preprint{APS/123-QED}

\title{Testing the Isotropy of the Universe with the CHIME/FRB Catalog I}

\author{Jun-Yi Shen}
\affiliation{Department of Astronomy, School of Physics, Huazhong University of Science and Technology, Wuhan 430074, China}

\author{Yuan-Chuan Zou}
\email{zouyc@hust.edu.cn}
\thanks{Corresponding author.}
\affiliation{Department of Astronomy, School of Physics, Huazhong University of Science and Technology, Wuhan 430074, China}

\date{\today}

\begin{abstract}
We test the isotropy of the Universe using 536 Fast Radio Bursts (FRBs) from the first CHIME/FRB catalog, employing two complementary statistical methods: the two-point angular correlation function (2PACF) and the sigma-map method, with observational biases corrected using CHIME injection data. Both analyses are consistent with the isotropic expectation, but are limited by the small sample size. The 2PACF suffers from an ill-conditioned covariance matrix, preventing robust parameter inference. The sigma-map angular power spectrum shows no statistically significant directional anisotropy, though the uncertainties at low multipoles remain large. 
Future FRB surveys are expected to improve these limitations. 
\end{abstract}

\maketitle


\section{Introduction}
The global shape of our Universe is a fundamental question, and various cosmological models have been investigated \cite{2003Natur.425..593L,1995PhR...254..135L}. Most researchers accept that the Universe is homogeneous and isotropic on large scales, a property known as the Cosmological Principle (CP) \cite{1993ppc..book.....P}, which is a fundamental assumption of modern cosmology. 
Based on the CP, the $\Lambda$CDM model describes a homogeneous and isotropic Universe on cosmological scales \cite{1922ZPhy...10..377F} and provides the theoretical framework for modern cosmology.

Various observations have tested the CP, among which the cosmic microwave background (CMB) temperature isotropy provides the most direct probe of large-scale isotropy, as measured by the Cosmic Background Explorer (COBE) \cite{1996ApJ...473..576F}, the Wilkinson Microwave Anisotropy Probe (WMAP) \cite{2011ApJS..192...18K}, and the Planck Collaboration \cite{2020A&A...641A...6P}. The large-scale structure of the Universe can also be studied using galaxy distributions from large-area surveys. Such surveys include optical observations such as the Sloan Digital Sky Survey (SDSS) \cite{2019MNRAS.483.2453S}. 
The large-scale radio sky has also been surveyed in the NRAO VLA Sky Survey (NVSS) \cite{2019JCAP...09..025B}, providing complementary data for isotropy tests. In addition, several approaches based on scalar observables have been employed to probe the anisotropy of the Universe, using tracers such as supernovae, gamma-ray bursts (GRBs), and quasars \cite{2015ApJ...808...39B, 2008ApJ...673..968B, 2021ApJ...908L..51S}. Overall, the CP is supported not only by observations, but also by its philosophical appeal and inherent mathematical symmetry.

However, recent studies have reported potential anisotropies on certain scales or in specific directions. For example, CMB anomalies have been reported by \cite{2020A&A...641A...6P}. Large-scale structure asymmetries have also been observed \cite{2004ApJ...605...14E}. In addition, a significant dipole anisotropy has been detected in the quasar distribution \cite{2021ApJ...908L..51S, 2022ApJ...937L..31S}. These findings suggest that the CP may not be exact in all contexts. Some large-scale structures (LSSs), such as the Giant Arc (GA), span overdense regions of $\sim$ 1 Gpc \cite{2022MNRAS.516.1557L}. The GRB sky distribution at z $\sim$ 2 shows small-angle deviations from isotropy \cite{2014A&A...561L..12H}. Many other observations also exist; see the review \cite{2022NewAR..9501659P, 2023CQGra..40i4001A}. 
These potential deviations from isotropy provide valuable insights into the fundamental properties and evolution of the Universe, and motivate alternative tests of the CP.

With the increasing number of detections, Fast Radio Bursts (FRBs) are emerging as a promising probe for testing large-scale isotropy. In this work, we use FRBs to study the CP. FRBs, first discovered by \cite{2007Sci...318..777L}, are highly luminous radio transients with millisecond durations. As the radio signal propagates, it interacts with free electrons along the line of sight, resulting in a dispersion measure (DM). The DM of FRBs contains a wealth of cosmological information, including constraints on the baryon fraction $f_b$ and the distribution of  “missing baryons” \cite{2020Natur.581..391M}.
FRBs are expected to become a powerful tool for studying cosmology \cite{2018ApJ...856...65W, 2020ApJ...903...83Z}. FRBs can serve as probes to estimate cosmological parameters, such as the Hubble constant \cite{2022MNRAS.515L...1W, 2022MNRAS.516.4862J}. More cosmological applications of FRBs are listed by a recent review \cite{2026enap....5..448G}.
FRBs offer several advantages for cosmological studies: they are extragalactic sources, wide-field surveys provide large samples, instrumental effects can be well calibrated, and high-quality catalogs are available for statistical analyses.

This paper is organized as follows. In Section~\ref{sec:data}, we describe the data selection. In Section~\ref{sec:method}, we present the methods. In Section~\ref{sec:result}, we show the results, and we conclude in Section~\ref{sec:conclusion}.

\section{CHIME Fast Radio Burst Data}\label{sec:data}

The Canadian Hydrogen Intensity Mapping Experiment (CHIME) was originally designed to detect the 21 cm neutral hydrogen emission to map the large-scale structure of the Universe \cite{2014SPIE.9145E..22B}.
For example, \cite{2024ApJ...963...23A} performed a cross-correlation of CHIME data with eBOSS Ly$\alpha$, and \cite{2025arXiv251119620C} reported the detection of the cosmological 21 cm intensity mapping signal. The CHIME sample is well suited for cosmological studies due to the availability of injection data, which can be used to correct for instrumental effects. Correcting for observational biases is essential, since uncorrected biases can lead to spurious results. 
For example, the study probing cosmic isotropy with gamma-ray bursts by \cite{2026JHEAp..5100549M} demonstrates that instrumental effects can introduce significant biases in the observed distribution.

Driven by its cosmological objectives, CHIME features a large collecting area and a wide field of view, operating in the frequency range 400–800 MHz, making it a powerful instrument for detecting FRBs. 
Moreover, as a wide-field telescope that performs continuous monitoring, CHIME surveys a large portion of the sky over extended periods. These properties allow for the building of a rich and statistically significant FRB catalog suitable for cosmological isotropy studies. \cite{2021ApJS..257...59C} presents the first CHIME FRB  catalog, containing 536 fast radio bursts from 2018 July 25 to 2019 July 1, including 62 bursts from 18 previously reported repeating sources\footnote{\url{https://chime-frb-open-data.github.io/}}. The latest catalog includes bursts observed between 25 July 2018 and 15 September 2023, totaling 3641 unique sources \cite{2026ApJS..283...34C}. Although 536 samples are still limited compared to galaxy surveys, the growing FRB sample allows for increasingly robust statistical analyses.

To study anisotropic features, simulated observations that preserve observational properties (e.g., sky exposure, resolution, and noise characteristics) are required.
During CHIME operations, synthetic calibration signals are used to monitor the telescope performance. They record information from antenna feeds, receivers, amplifiers, and digital systems, including signal strength, noise levels, and system gain. The data used in this work are publicly available at \url{https://chime-frb-open-data.github.io/injections/}.
These synthetic datasets match the real data in terms of sky coverage and are used to correct for anisotropic effects introduced by the instrument.  In this work, our analysis is carried out using CHIME Catalog 1. The recently released CHIME Catalog 2 contains more samples with smaller fluctuations; however, the corresponding injection data for Catalog 2 are not yet publicly available and therefore were not used in our analysis \cite{2026ApJS..283...34C}.

\section{Method}\label{sec:method}
We adopt two complementary approaches to quantify anisotropy: the two-point angular correlation function (2PACF) and the sigma-map method. The 2PACF is a standard statistical estimator that is used to quantify the angular clustering of sources on the celestial sphere. Under the assumption of statistical isotropy, the 2PACF depends only on the angular separation $\theta$. To test this assumption, we compare the 2PACF measured from the observational data with that derived from the null hypothesis condition. Deviations between the two provide evidence for possible departures from isotropy. To further investigate possible directional dependence, the 2PACF can be generalized to include sky direction dependence $\omega=\omega(\theta, \mathrm{RA}, \mathrm{Dec})$. We construct a sigma-map by computing the variance of 2PACF evaluated locally within spherical caps covering the sky. This procedure yields a direction-dependent scalar field that encodes spatial variations of the clustering properties. Finally, we perform a spherical harmonic decomposition of the sigma-map to characterize the angular scale of the anisotropy through its power spectrum $C_{\ell}$.
\subsection{Two point angular correlation function}
We denote the local matter density by $\rho(\mathbf{x})$. 
The large-scale structure of the Universe can be described by the density perturbation field
\begin{equation}\label{Eq:delta}
    \delta(\mathbf{x}) = \frac{\rho(\mathbf{x}) - \bar{\rho}}{\bar{\rho}}.
\end{equation}
The large-scale density field of the Universe is well approximated as a Gaussian random field \cite{2020A&A...641A...6P}.
Although certain early-Universe processes, such as topological defects, can generate non-Gaussian perturbations, these contributions are generally subdominant. 
On large scales, where the evolution remains in the linear regime, the density field is well approximated as a Gaussian.

In this work, we focus on large-scale statistical properties, where the field is expected to be approximately statistically isotropic. 
Non-Gaussian effects arising from small-scale nonlinear evolution have a negligible impact on our analysis.
All statistical properties are fully determined by two-point statistics (i.e., the two-point correlation function or equivalently the power spectrum; see, e.g., Chapter 6.1.3 \cite{2010gfe..book.....M}).
Therefore, the clustering properties can be fully characterized by the two-point correlation function, defined as:
\begin{equation} \label{xidelta}
    \xi(\mathbf{r}) = \langle \delta(\mathbf{x}) \, \delta(\mathbf{x} + \mathbf{r}) \rangle,
\end{equation}
where $\langle \cdot \rangle$ denotes the ensemble average and $\mathbf{r}$ is the separation vector.
For a Gaussian random field, all connected $n$-point correlation functions vanish for $n>2$.
The three-dimensional two-point correlation function of galaxies can be estimated from Eq.~\ref{xidelta} by the natural estimator 
\begin{equation}
\xi(r) = \frac{DD(r) -  RR(r)}{RR(r)}.
\end{equation}
An improved estimator was proposed by \cite{1993ApJ...412...64L}:
\begin{equation}
\xi(r) = \frac{DD(r) - 2\,DR(r) + RR(r)}{RR(r)},
\end{equation}
where $DD(r)$ is the observed number of galaxy–galaxy pairs with separations in the range $r \pm \Delta r/2$. $RR(r)$ is the number of random–random pairs drawn from a uniform (Poisson) distribution, and $DR(r)$ is the number of cross-pairs between the real and random samples. The two-point correlation functions method has been widely used in large-scale structure (LSS) studies, such as \cite{2024MNRAS.527.7400F,2025ApJ...993..133F}.

The CHIME catalog primarily provides the sky positions of FRBs. 
Since only a limited number of FRBs have measured redshifts, most sources provide only angular information, i.e., right ascension (RA) and declination (Dec).
In this case, the angular correlation function represents a line-of-sight projection of the underlying three-dimensional clustering, where contributions from different distances are effectively integrated together.
For such a two-dimensional dataset, we estimate the 2PACF \cite{1993ApJ...412...64L}:
\begin{equation}\label{2pacf}
    \omega(\theta) = \frac{DD(\theta) - 2\,DR(\theta) + RR(\theta)}{RR(\theta)}.
\end{equation}
Here, $DD(\theta)$, $DR(\theta)$, and $RR(\theta)$ denote the normalized data–data, data–random, and random–random pair counts at angular separation $\theta$, respectively.

\subsection{Error estimation for 2PACF}
The observed $\omega$ provides an estimate of the true underlying cosmic $\omega$, but is affected by the survey geometry and statistical fluctuations in pair counts.  In principle, the uncertainty can be estimated from the variance of an ensemble of realizations. However, we can observe only a single realization of the Universe.
We can either use mock samples to estimate the uncertainty or apply the jackknife method as an internal error estimator.
In this work, we employ both methods, the mock catalog and the jackknife, to validate our error estimates.
\cite{2009MNRAS.396...19N} compared these two approaches and found that mock catalogs generally provide more reliable statistical uncertainties.
In contrast, the jackknife method can systematically deviate from the true uncertainty compared to the mock catalog results. As described in \cite{2022MNRAS.514.1289M}, certain techniques can reduce the bias of the jackknife estimates. In this work, we adopt the mock catalog to estimate the uncertainty of $\omega$, with the jackknife method serving as a supplementary check.
\subsubsection{Mock catalog} \label{sec:Mock}
The mock catalog is designed to reproduce the statistical properties of the observed data. The covariance matrix is calculated as
\begin{equation}
\text{Cov}_{ij} = 
\frac{1}{N_{\rm{mock}}-1} 
\sum_{k=1}^{N_{\rm{mock}}} 
\left[ \omega_k(\theta_i) - \bar{\omega}(\theta_i) \right]
\left[ \omega_k(\theta_j) - \bar{\omega}(\theta_j) \right],
\end{equation} 
where \(i,j = 1,2,\dots,N_b\) label the angular bins, 
\(\omega_k(\theta_i)\) is the 2PACF measured from the \(k\)-th mock realization, 
and \(\bar{\omega}(\theta_i)\) denotes the ensemble average over all mock realizations.  The uncertainty in the 
2PACF is $ \Delta \omega(\theta_i)
=\sqrt{\mathrm{Cov}_{ii}} $.
Mock catalogs can be regarded as different realizations of the Universe. They provide an ensemble of synthetic replicas of real observations that mimic the underlying probability distribution of the observed data. We use a lognormal model to generate mock catalogs. The lognormal distribution naturally arises in stochastic processes where a variable is formed as the product of many independent random factors. It provides a good description of non-linear small-scale regions, while approaching a Gaussian distribution in the linear regime, thereby ensuring consistency with linear theory \cite{1991MNRAS.248....1C}. For example, \cite{2017MNRAS.466.1444C} demonstrated that the lognormal model can successfully describe the distribution of cosmological matter. We used a cosmological simulation of the lognormal matter distribution provided in \cite{2013JCAP...12..030C}. The cosmological parameters are listed in Table~\ref{Table:mockpara}, mostly taken from \cite{2020A&A...641A...6P}. The mock catalog is constructed from a simulation snapshot at redshift $z=0.5$. Although FRBs are expected to span a wide redshift range, our analysis focuses on angular statistics, which are primarily sensitive to the projected distribution on the sky. Therefore, using a single snapshot provides a reasonable approximation for testing isotropy.
We verified that the estimated uncertainties remain consistent when using mock catalogs constructed from different snapshot redshifts, 
indicating that our conclusions are not sensitive to the specific choice of snapshot.

\begin{table}[h]
\centering
\caption{Parameters used for the lognormal mock catalog to estimate the 2PACF errors.}
\label{Table:mockpara}
\begin{tabular}{l c l}
\hline\hline
Parameter & Value & Description \\
\hline
$z$ & 0.5 & Redshift of the snapshot \\
$L_x, L_y, L_z$ & 530 Mpc/$h$ & Box size in each dimension \\
$N_{\rm gal}$ & $\sim 600$ & Number of FRB hosts \\
$b$ & 2.0 & Linear bias \\
$\Omega_c h^2$ & 0.1198 & Cold dark matter density \\
$\Omega_b h^2$ & 0.02225 & Baryon density \\
$h$ & 0.67021 & Hubble parameter ($H_0/100$) \\

\hline
$N_{\rm real}$ & 200 & Number of lognormal realizations \\
$P_{\rm nmax}$ & 128 & Grid size per dimension \\
\hline
\end{tabular}
\end{table}

\begin{figure}[htbp]  
    \centering
    \includegraphics[width=0.4\textwidth]{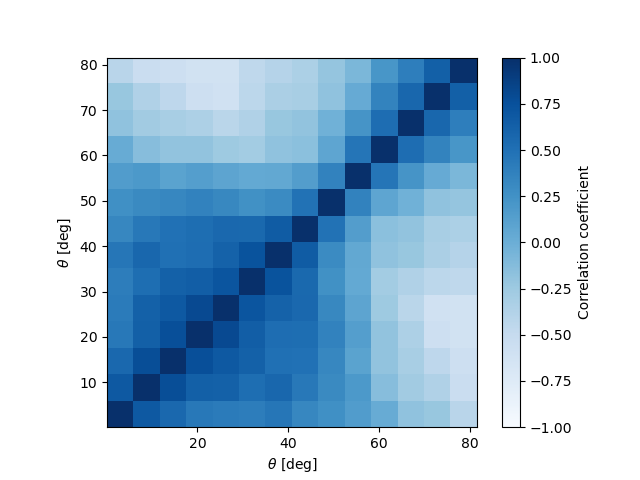}  
    \caption{Normalized correlation matrix of the 2PACF, defined as $\mathrm{Corr}_{ij} = \mathrm{Cov}_{ij}/\sqrt{\mathrm{Cov}_{ii}\mathrm{Cov}_{jj}}$. The 2PACF analysis is performed over the angular range $0^\circ \leq \theta \leq 80^\circ$.}
    \label{fig:corv}  
\end{figure}

\subsubsection{Jackknife} 
The jackknife method estimates uncertainties directly from the data itself, without relying on simulations or mock catalogs. As an internal resampling technique, it provides a model-independent way to estimate the covariance matrix for clustering statistics from a single realization of the data. The jackknife method yields covariance estimates comparable to those of ensemble methods with fewer realizations \cite{2016arXiv160600233E}, and the choice of jackknife sub-sample scale does not significantly bias large-scale structure error estimates \cite{2021MNRAS.505.5833F}.

In this approach, the full catalog of 536 FRBs is divided into roughly equal subsets $N_{\rm JK}=20$. For each jackknife sample, one subset is omitted and the 2PACF is calculated using the remaining subsets. Denoting the 2PACF measured in the $i$-th jackknife sample as $\omega_i(\theta)$, the covariance matrix is estimated as
\begin{equation}
C_{jk}(\theta, \theta') = \frac{N_{\rm JK}-1}{N_{\rm JK}}
\sum_{i=1}^{N_{\rm JK}} \left[ \omega_i(\theta) - \bar{\omega}(\theta) \right]
\left[ \omega_i(\theta') - \bar{\omega}(\theta') \right],
\end{equation}
where $\bar{\omega}(\theta) = \frac{1}{N_{\rm JK}} \sum_{i=1}^{N_{\rm JK}} \omega_i(\theta)$ is the mean over all jackknife samples. The diagonal elements of $C_{jk}$ represent the variance of $\omega(\theta)$ in each angular bin, which we adopt as the $1\sigma$ uncertainty of the 2PACF. 

\subsection{Sigma-map} \label{sec:sigmap}
One of the key assumptions of the CP is statistical isotropy (SI).
SI states that the statistical properties of the Universe are invariant under rotations.
This implies that there is no preferred direction in the sky. 
The 2PACF is a direction-independent (isotropic) quantity, depends only on the angular separation between pairs of sources, and therefore does not retain directional information. 
The sigma-map method extends this idea by constructing a direction-dependent field from the spatial variation of the local 2PACF \cite{2017MNRAS.466.2799B, 2007A&A...464..479B}. In this approach, the sky is divided into different regions, and the 2PACF is computed locally within each region. The sigma-map is then defined as a scalar field that quantifies the deviation of the local 2PACF from the isotropic expectation.
The statistical measure associated with the sigma-map is given by the variance of these local deviations, which allows us to probe possible directional dependence of clustering statistics on the celestial sphere.

The celestial sphere $S^2$ is divided into $N_{\rm cap}$ spherical caps $\Omega_j(RA_j, Dec_j ,\gamma_0)$, each defining a local region centered in a specific direction. The $\gamma_0$ denotes the aperture of the cap.
For each region $\Omega_j$, we define a scalar quantity $\sigma_j$ that quantifies the deviation of the local 2PACF from statistical isotropy. This quantity is given by the variance of $\Delta_j(\gamma_i; \gamma_0)$ over the angular separation bins:
\begin{equation}
\sigma_j^2 \equiv \frac{1}{N_{\rm bins}} \sum_{i=1}^{N_{\rm bins}} \Delta_j^2(\gamma_i; \gamma_0),
\end{equation}
where $\Delta_j(\gamma_i; \gamma_0)$ represents the difference between the normalized pair counts in the $j$-th cap and the expectation of a statistically isotropic distribution, evaluated for angular separations within the interval $(\gamma_i - 0.5\delta, \gamma_i + 0.5\delta]$, with $i = 1, \ldots, N_{\rm bins}$.
Here, the $j$-th cap is centered in a specific direction $(RA_j, Dec_j)$ on the sky, so that $$\sigma_j \equiv \sigma(\hat{n}_j), \ \hat{n}_j \equiv \hat{n}(\mathrm{RA}_j, \mathrm{Dec}_j).$$
The set of caps $\sigma_j$ provides a nearly full-sky sampling of the angular correlation properties.
\section{Result}\label{sec:result}
In this section, we present our analysis of the FRB catalog. Each repeating source is counted only once in the statistical analysis. 
Bursts toward the inner Galaxy are excluded, since they lie at low Galactic latitudes and may introduce bias. The angular resolution of CHIME is approximately $0.3^\circ$, so the $\omega(\theta)$ measured at small angular separations ($\theta < 1^\circ$) is not reliable. As described in \cite{2025arXiv251113931B}, the information at small angles is smeared by the Gaussian beam window, reducing the reliability of the correlation function in this region, whereas at larger angular scales the correlation function can be reliably measured. Since our goal is to search for large-scale anisotropy, the beam smoothing of small-scale fluctuations has a negligible impact on our analysis. We proceed to show the measurements of the 2PACF and the angular power spectrum as follows.
\subsection{2PACF result}
The 2PACF is defined in Eq.~\ref{2pacf} and is computed with the \texttt{TreeCorr} package \footnote{\url{https://github.com/rmjarvis/TreeCorr}}. Data–data pair counts, DD($\theta$), are calculated by considering all unique pairs of FRB events in the CHIME/FRB catalog. For each pair, the angular separation is calculated and assigned to the corresponding angular bin $\theta \pm \Delta \theta/2$ of DD($\theta$). To estimate the correlation function using the Landy–Szalay estimator, both the DD and the RR pair counts are required. The RR pairs are generated in the same way as DD, but using a synthetic catalog that follows the observational properties of the survey. This random catalog is constructed to account for instrumental and observational effects, such as the sky exposure and beam response of CHIME, as well as selection effects. Specifically, the probability of selecting a random point at a given position (RA, Dec) is proportional to the sky exposure, i.e., 
\begin{equation}
    P(\mathrm{RA}, \mathrm{Dec}) \propto \mathrm{exposure}(\mathrm{RA}, \mathrm{Dec}).
\end{equation}
The use of RR in the estimator corrects for these non-uniformities, ensuring that the measured 2PACF reflects intrinsic clustering rather than observational biases.
The exposure time and pointing direction of CHIME are recorded for all observations, allowing the construction of an exposure map. Using this map, a synthetic RR catalog can be generated that reflects the non-uniform sky coverage. In addition, CHIME provides injection data, as described in Section~4 of \cite{2021ApJS..257...59C}, which contain simulated bursts injected into the real-time pipeline to characterize the completeness of detection and selection effects. These injection datasets are commonly used alongside the CHIME/FRB catalog to correct for observational biases. Both the real-time injection data and the catalog presented in \cite{2024ApJ...963...23A} serve this purpose, allowing a more accurate construction of the RR catalog.

\begin{figure}[htbp]  
    \centering
    \includegraphics[width=0.4\textwidth]{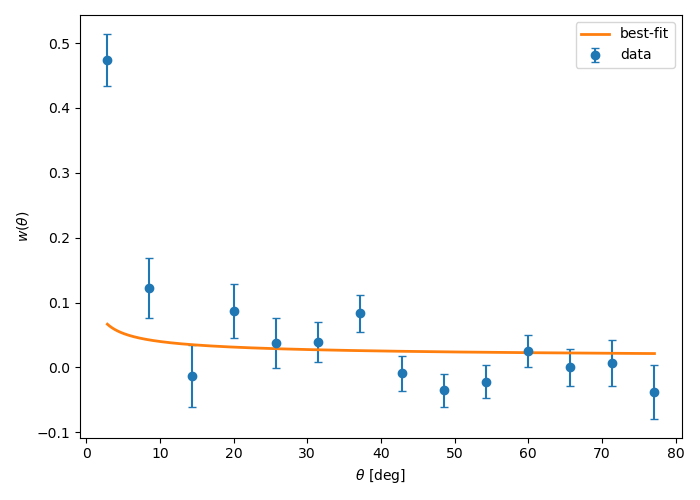}  
    \caption{Mosaic of the 2PACF for the FRB angular distribution derived from the first CHIME/FRB catalog. The 2PACF is computed using the Landy--Szalay estimator over the angular range $0^\circ \leq \theta \leq 80^\circ$. The orange line is the power-law best fit of $\omega(\theta)$. The best-fit parameters are $A_w = 0.0958$ and $\gamma = 1.536$.   }
    \label{fig:2pacf}  
\end{figure}

To estimate statistical uncertainties, we generate $N_{\rm mock} = 200$ mock catalogs representing independent realizations of the Universe. These mock catalogs are different from the RR catalog, as their purpose is solely to estimate the covariance of the two-point correlation function. For each mock, the RR data (CHIME injection) is used in the calculation of the 2PACF. The resulting covariance matrix is shown in Fig.~\ref{fig:corv}. 
Under the assumptions of isotropy and homogeneity, the 2PACF on large-scale is expected to follow a power-law form \cite{1980lssu.book.....P}:
\begin{equation}
    \omega(\theta) = A_w \, \theta^{1-\gamma}.
\end{equation}
The $A_w$ represents the amplitude and $1-\gamma$ is the power-law index. This parametrization is used to test whether the Universe is isotropic. In this work, only the large-angle ($\theta > 1^\circ$) $\omega(\theta)$ measurements are considered, where the effect of the CHIME beam is negligible. The angular bins are defined to be linearly spaced over the range 
$0^\circ < \theta < 80^\circ$, using 14 bins. 
A sufficiently large ratio between the number of mock realizations and the number of bins is required to obtain a stable covariance estimate and its inverse \cite{2007A&A...464..399H}. In our case, with 14 bins and 200 mocks, this condition is reasonably satisfied. Although the number of mock realizations is modest, 
it remains larger than the number of angular bins and is sufficient for the present exploratory analysis.
The 2PACF measured from the CHIME catalog, together with the error bars derived from the mock catalogs, is shown in Fig.~\ref{fig:2pacf}. To further validate the reliability of the error estimation, we also computed jackknife errors and compared them with mock-based uncertainties. The results are summarized in Table~\ref{tab:jackmock}. The jackknife uncertainties are generally larger than those estimated from the mock catalogs by factors of order \(1.5\text{--}2\). These results suggest that the statistical uncertainties are difficult to estimate robustly with the current FRB sample.

We use \texttt{emcee} \cite{2013PASP..125..306F} to perform a Bayesian parameter estimation. 
When fitting the model, we exclude measurements at angular scales $\theta \leq 1^\circ$, 
since these scales are affected by the angular resolution of CHIME ($\sim 0.3^\circ$). 
The parameters are constrained by minimizing
\begin{equation}
\chi^2(\gamma, A_w)
=
[\omega(\theta)-\omega_{\rm model}(\theta;\gamma,A_w)]^T
\mathrm{Cov}^{-1}
[\omega(\theta)-\omega_{\rm model}(\theta;\gamma,A_w)],
\end{equation}
where $\mathrm{Cov}$ is the mock-based covariance matrix described in Section~\ref{sec:Mock}. This form properly accounts for correlations between different angular bins. We adopt uniform priors for the parameters:
\[
10^{-6} < A_w < 10^{-1}, \qquad 0.5 < \gamma < 3.
\] We run the MCMC sampler for $10^4$ steps, and confirm that the chains have converged.

\begin{table}[h]
\centering
\caption{Comparison between jackknife and mock uncertainties.}
\label{tab:jackmock}
\begin{tabular}{cccc}
\hline
bin index & $\sigma_{\rm jack}$ & $\sigma_{\rm mock}$ & $\sigma_{\rm jack}/\sigma_{\rm mock}$ \\
\hline
1  & 0.2247 & 0.04339 & 5.18 \\
2  & 0.0967 & 0.05016 & 1.93 \\
3  & 0.0922 & 0.05174 & 1.78 \\
4  & 0.0406 & 0.04535 & 0.90 \\
5  & 0.0315 & 0.04159 & 0.76 \\
6  & 0.0587 & 0.03380 & 1.74 \\
7  & 0.0801 & 0.03102 & 2.58 \\
8  & 0.0486 & 0.02992 & 1.62 \\
9  & 0.0428 & 0.02766 & 1.55 \\
10 & 0.0476 & 0.02782 & 1.71 \\
11 & 0.0571 & 0.02716 & 2.10 \\
12 & 0.0547 & 0.03101 & 1.76 \\
13 & 0.0659 & 0.03846 & 1.71 \\
14 & 0.0688 & 0.04467 & 1.54 \\
\hline
\end{tabular}
\end{table}

The parameter estimation results are shown in Fig.~\ref{fig:2pacf}. The best-fit parameters are
\[
A_w = 0.0958^{+0.0032}_{-0.0071},
\qquad
\gamma = 1.536^{+0.162}_{-0.129}.
\]
The measured 2PACF is affected by substantial statistical uncertainties, and the inferred goodness-of-fit depends sensitively on the adopted covariance estimator. Using the mock-based covariance matrix, we obtain a reduced $\chi^2 \sim 16$ at the best-fit parameters.
Although the fitted model reproduces the overall trend of the measured 2PACF, the reduced $\chi^2$ obtained using the mock-based covariance matrix is relatively large, suggesting an apparent tension between the model and the data. 
However, this large reduced $\chi^2$ may be driven by uncertainties in the covariance estimation rather than a true discrepancy between the model and the data. The covariance matrix is ill-conditioned, indicating that the current statistical inference is covariance-limited.
This behavior is further supported by the eigenvalue spectrum of the covariance matrix, which shows a rapid decay with only a few dominant modes. As a consequence, the $\chi^2$ statistic is not uniformly constrained by all angular bins, but is instead effectively dominated by a small number of leading eigenmodes.
This arises from the highly correlated structure of the 2PACF estimator and the sparse sampling of the FRB catalog. In this regime, the nominal number of angular bins significantly overestimates the true statistical dimensionality of the data.
Robust covariance estimation remains challenging for the current FRB sample, and therefore the statistical significance of any deviation from the power-law model should be interpreted with caution.

\subsection{Sigma-map angular power spectrum}

We adopt the same parameters as in \cite{2017MNRAS.466.2799B}: $\gamma_0 = 90^\circ$ as the reference angle, $N_{\rm bins} = 49152$ corresponding to a HEALPix resolution of $N_{\rm side} = 64$, and $j = 1, \dots, N_{\rm caps}=768$ enumerates all the spherical caps that cover the sky. This choice ensures sufficient angular resolution for each cap while maintaining statistical robustness in the estimation of $\sigma_j^2$. Here, $\Delta_j(\gamma_i; \gamma_0)$ is the difference normalized by the total number of pairs in the angular bin $\gamma_i$ relative to the reference angle $\gamma_0$, and $N_{\rm bins}$ is the total number of angular bins considered in each cap.

The sigma-map $\sigma(\hat{n})$ can be expanded in spherical harmonics as
\begin{equation}
\sigma(\hat{n}) = \sum_{\ell=0}^{\ell_{\rm max}} \sum_{m=-\ell}^{\ell} A_{\ell m} Y_{\ell m}(\hat{n}),
\end{equation}
where $A_{\ell m}$ are the spherical harmonic coefficients. The corresponding angular power spectrum is defined by
\begin{equation}
C_\ell \equiv \frac{1}{2\ell+1} \sum_{m=-\ell}^{\ell} |A_{\ell m}|^2,
\end{equation}
which quantifies the variance of $\sigma$ in each multipole $\ell$. The $C_{\ell}$ are generated using the HEALPix \citep{2005ApJ...622..759G} through its Python implementation healpy
\citep{2019JOSS....4.1298Z}. The angular power spectrum $C_{\ell}$ of the sigma-map quantifies the amplitude of anisotropy as a function of the angular scale. The higher values of $C_{\ell}$ indicate stronger spatial variations of the local clustering signal at the corresponding multipole $\ell$.

\begin{figure}[htbp]  
    \centering
    \includegraphics[width=0.4\textwidth]{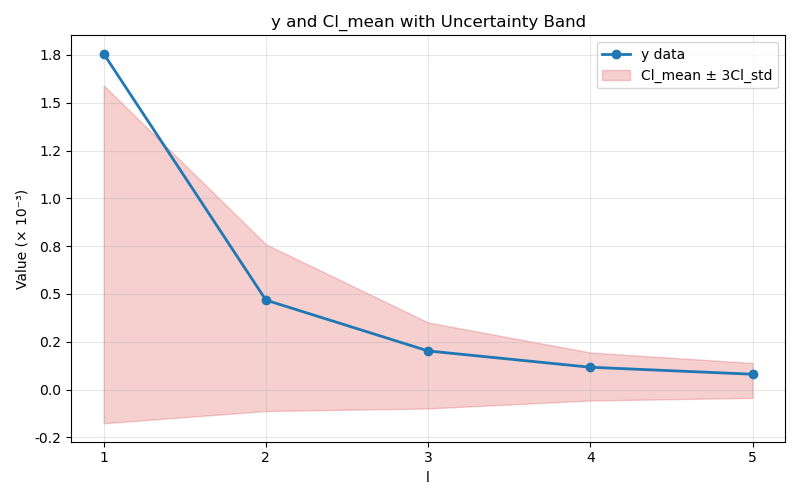}  
    \caption{ The shaded region represents the $\pm 3\sigma$ confidence interval obtained from 100 Monte Carlo realizations, in which sources are randomly drawn from the injection catalog. We note that the negative values of $C_\ell$ are produced by statistical fluctuations (i.e., arise from the $\pm\sigma$ scatter) and do not have physical meaning. The solid blue line shows the $C_\ell$ computed from the isotropic mock catalog. }
    \label{fig:sigmap}  
\end{figure}

The MC results indicate that the angular power spectrum operates in a low signal-to-noise regime, making it difficult to distinguish between genuine signal and statistical fluctuations.
We define $C_\ell^{\mathrm{mean}}$ as the ensemble average over all MC realizations, and $C_\ell^{\mathrm{std}}$ as the corresponding standard deviation, which characterizes the statistical uncertainty arising from sample variance and shot noise.
We find that the ratio $C_\ell^{\mathrm{mean}} / C_\ell^{\mathrm{std}} \sim 1-2$ remains small across all multipoles, confirming that the measurements are dominated by statistical fluctuations.
Overall, the angular power spectrum is fluctuation-dominated, and the current data do not allow a statistically robust inference of any underlying signal.

\section{Conclusion and discussion}\label{sec:conclusion}

We use the 2PACF and the sigma-map method to test cosmic isotropy in the FRB sky distribution. The 2PACF provides a global statistical measure of angular clustering, while the sigma-map probes possible directional dependence by quantifying spatial variations of the local 2PACF across the sky. The data are taken from the first CHIME/FRB catalog, containing 536 FRB events sparsely distributed across the sky.

The current FRB sample does not allow a statistically definitive test of cosmic isotropy. We can only conclude that no statistically significant evidence for anisotropy is found, although the constraining power remains limited by sample variance and shot noise. The 2PACF measurements are affected by Poisson shot noise due to the limited number of FRB pairs in each angular bin, with uncertainties approximately scaling as $\sigma_w \sim 1/\sqrt{DD}$.
In addition to shot noise, the covariance structure is highly non-trivial, with strong inter-bin correlations leading to a reduced effective number of degrees of freedom. As a result, the inferred clustering signal is dominated by statistical fluctuations in a low-dimensional subspace of the data vector, rather than by independent constraints from all angular bins.
Consequently, the goodness-of-fit becomes sensitive to the covariance estimation method. In particular, the mock-based covariance matrix yields a large reduced $\chi^2$, which reflects the combination of limited statistical power and the low-rank nature of the covariance rather than a robust detection of a model discrepancy.

The sigma-map analysis is limited by large statistical uncertainties and does not allow meaningful constraints on large-scale anisotropy. The angular power spectrum $C_\ell$ decreases monotonically from $\ell=1$ to $\ell=5$, and all multipoles are consistent with isotropy within the $3\sigma$ uncertainty range. However, the uncertainties remain large at low multipoles, particularly at $\ell=1$, preventing strong constraints on possible large-scale anisotropy. 
Therefore, no statistically significant evidence for anisotropy is found, although the constraining power remains limited by sample variance and shot noise.

The main limitation of the present analysis is the small sample size, which leads to sparse pair counts and shot-noise-dominated statistics. In addition, the covariance matrix is highly ill-conditioned, reflecting strong bin-to-bin correlations and a reduced effective rank, which further limits the reliability of parameter inference. A secondary systematic effect arises from the clustering of FRB host galaxies, which are preferentially associated with star-forming regions \cite{2024Natur.635...61S}, and may introduce additional small-scale correlations, although this effect is subdominant at the large angular scales considered here.

Future CHIME/FRB catalogs are expected to substantially enhance the statistical power of isotropy tests. In particular, Catalog 2 \citep{2026ApJS..283...34C} and the forthcoming Catalog 3 already contain thousands of FRBs, offering an order-of-magnitude increase in sample size compared to the current analysis. Once the corresponding injection datasets become publicly available, more accurate and complete corrections for observational biases (such as exposure, completeness, and beam effects) will be possible. These improvements will not only reduce shot noise and stabilize the covariance estimation in the 2PACF analysis, but also enable tighter constraints on the sigma-map power spectrum at low multipoles. With larger samples and refined bias corrections, FRBs will provide increasingly robust tests of the Cosmological Principle, potentially reaching the sensitivity required to detect or rule out subtle large-scale anisotropies.

Future FRB surveys are expected to significantly improve these limitations. An increase in the sample size by at least an order of magnitude from facilities such as CHIME/FRB and ASKAP will substantially reduce statistical uncertainties and improve covariance stability. In particular, the arcsecond-scale localization capability provided by CHIME/FRB together with the Outriggers VLBI system \cite{2025ApJ...993...55C} will enable precise studies of FRB host galaxies, redshifts, and progenitor populations. With larger and better-localized samples, FRBs will provide significantly improved constraints on cosmic isotropy and the large-scale structure of the Universe.

\begin{acknowledgments}
This work was supported by the National Natural Science Foundation of China (grant No.
12041306).
\end{acknowledgments}

\appendix

\bibliography{apssamp}

\end{document}